\documentclass[authoryear,preprint,11pt,a4paper]{elsarticle}

\usepackage{color,soul}
\usepackage{graphics}
\usepackage{epsfig}
\usepackage{rotating}
\usepackage{amssymb}

\usepackage{supertabular}

\usepackage{longtable}

\usepackage{lscape}

\usepackage{fullpage}

\journal{Icarus}

\begin{document}

\begin{frontmatter}



\title{Surface composition and taxonomic classification of a group of near-Earth and Mars-crossing asteroids}

\author[IFP]{Juan A. Sanchez\corref{cor1}}
\ead{sanchez@mps.mpg.de}

\author[]{Ren\'{e} Michelsen\fnref{fn1}}
\author[UND]{Vishnu Reddy\fnref{fn2}}
\author[MPS]{Andreas Nathues}

\address[IFP]{Institut f$\ddot{u}$r Planetologie, Westf\"alische Wilhelms-Universit\"at M$\ddot{u}$nster, Germany}
\address[UND]{Department of Space Studies, University of North Dakota, Grand Forks, ND 58202, USA}
\address[MPS]{Max Planck Institut f$\ddot{u}$r Sonnensystemforschung, Max Planck Str.2, 37191 Katlenburg-Lindau, Germany}

\fntext[fn1]{Visiting Scientist at the Max Planck Institut f$\ddot{u}$r Sonnensystemforschung.}

\fntext[fn2]{Visiting Astronomer at the Infrared Telescope Facility, which is operated by the University of Hawaii under Cooperative Agreement No. NNX-08AE38A with the National Aeronautics and Space Administration, Science Mission Directorate, Planetary Astronomy Program.}

\cortext[cor1]{Corresponding author at: Max Planck Institut f$\ddot{u}$r Sonnensystemforschung, Max Planck Str.2, 37191 Katlenburg-Lindau, Germany.}


\begin{abstract}

In the past, constraining the surface composition of near-Earth asteroids (NEAs) has been difficult due to the lack of high quality near-IR spectral data (0.7-2.5 $\mu$m) that contain mineralogically diagnostic absorption 
bands. Here we present visible ($\sim$ 0.43-0.95 $\mu$m) and near-infrared ($\sim$ 0.7-2.5 $\mu$m) spectra of nine NEAs and five Mars-crossing asteroids (MCs). The studied NEAs are: 4055 Magellan, 
19764 (2000 NF5), 89830 (2002 CE), 138404 (2000 HA24), 143381 (2003 BC21), 159609 (2002 AQ3), 164121 (2003 YT1),  241662 (2000 KO44) and  2007 ML13. The studied MCs are: 1656 Suomi, 2577 Litva, 
5407 (1992 AX), 22449 Ottijeff and 47035 (1998 WS). The observations were conducted with the NTT at La Silla, Chile, the 2.2 m telescope at Calar Alto, Spain, and the IRTF on Mauna Kea, Hawai'i. The 
taxonomic classification (Bus system) of asteroids showed that all observed MC asteroids belong to the S-complex, including the S, Sr and Sl classes. Seven of the NEAs belong to the S-complex, including the S, Sa, 
Sk and Sl classes, and two NEAs were classified as V-types. The classification of the NEA 164121 (2003 YT1) as a V-type was made on the basis of its near-infrared spectrum since no visible spectrum is available for this 
asteroid. A mineralogical analysis was performed on six of the asteroids (those for which near-IR spectra were obtained or previously available). We found that three asteroids (241662 (2000 KO44), 
19764 (2000 NF5), 138404 (2000 HA24)) have mafic silicate compositions consistent with ordinary chondrites, while three others (4055 Magellan, 164121 (2003 YT1), 5407 (1992 AX)) are pyroxene-dominated 
basaltic achondrite assemblages. In the case of 5407 (1992 AX) we found that its basaltic surface composition contrasts its taxonomic classification as a S-type.

\end{abstract}

\begin{keyword}

Asteroids \sep Near-Earth objects \sep Spectroscopy \sep Infrared observations \sep Meteorites



\end{keyword}

\end{frontmatter}



\section{Introduction}

Among the small body population, near-Earth asteroids (NEAs) are defined as asteroids having perihelion distances of $\le$ 1.3 AU. The NEA population shows a great 
diversity in terms of taxonomic classification, including almost all classes of asteroids found in the main belt \citep{2002aste.conf..255B}. However, of all taxonomic types, S-complex asteroids are the 
dominant class among NEAs. This high proportion of S-complex is caused in part by a selection effect, due to the fact that S-complex asteroids have higher albedos than C-types (the dominant class among 
all main-belt asteroids), making them easier to discover. In addition, there are some regions in the main belt that contribute more than others to the delivery of objects into the inner solar system. According to some 
dynamical models \citep[e.g.,][]{1999Icar..139..295M, 2000Sci...288.2190B, 2002aste.conf..395B, 2002aste.conf..409M}, the 3:1 mean-motion resonance with Jupiter (occurring at $\sim$ 2.5 AU), and the 
$\nu_{6}$ secular resonance at the inner asteroid belt (where S-complex are most common) provide the most significant fraction of objects to the near-Earth population.   


Based on their orbital parameters NEAs are divided into three subgroups called Amor, Apollo, and Aten asteroids \citep{1979aste.book..253S}. Some of these objects are also classified as potentially hazardous 
asteroids (PHAs), which are defined as objects whose minimal orbital intersection distance (MOID) with Earth is $\le$ 0.05 AU. In addition to this dynamical classification, there is another group of objects called 
Mars-crossing asteroids (MCs), whose orbits cross or approach that of Mars, and are characterized by perihelion distances 1.30 $<$ q $<$ 1.66 AU. Similarly to NEAs, the MC population is dominated by 
S-complex asteroids \citep{2002A&A...391..757A, 2004Icar..170..259B}. With time, MCs can eventually become NEAs as a result of close approaches with Mars, that move them into a resonance, and from there, to 
the near-Earth space \citep{2000Icar..145..332M}. 

NEAs are of great interest for study, and due to their proximity they could be the source of some of the meteorites found on Earth. Thus, establishing links between the spectral properties of NEAs and those measured in 
the laboratory for meteorite samples is fundamental to determine their possible parent bodies \citep{2002aste.conf..653B, 2002aste.conf..255B}. Recent studies show that about 2/3 of the observed NEAs have surface 
compositions consistent with LL chondrites, which represent only about 10\% of the ordinary chondrites falls \citep{2008Natur.454..858V,2010A&A...517A..23D,2013Icar..222..273D}. These results contradict 
the general assumption that the spectral characteristics of the most common type of NEAs should be consistent with those of the most common meteorites to fall on Earth. The reason for this discrepancy still remains 
unclear.

NEAs are also interesting for study because they are relatively easy to reach compared to their counterparts in the main belt. Some NEAs have low velocities with respect to Earth ($\Delta$V), which make them especially 
accessible for sample-return and manned missions \citep[e.g.,][]{2000M&PSA..35R.145S, 2009M&PS...44.1825A, 2012Icar..221..678R}. Furthermore, it is now recognized the threat that some NEAs could represent 
for civilization in the eventuality of collision \citep[e.g.,][]{1994Natur.367...33C, 2002aste.conf..739M}. Hence, having a better understanding of their physical properties could help us to develop mitigation strategies in 
the future.  

In this work we report visible (VIS) and near-infrared (NIR) spectroscopic observations of five MCs and nine NEAs. Although limited, our sample is diverse and comprises both basaltic an ordinary chondrite-like 
asteroids. Furthermore, three objects are classified as PHAs and one object is considered as a possible target for robotic and manned missions. Thus, the present work seeks to obtain new compositional information that 
could help tracing the origin of some meteorites, contribute in planning future robotic and manned missions, and contribute in the development of mitigation strategies. In addition, one of the goals of this paper is to 
study the compositional trend in the MC population and compare it with the NEA population. This will help us understand the source regions from which the NEAs are derived, and if the same compositional types 
dominate both populations. This is a preliminary study since our current sample is small. The selection of these particular asteroids for study was influenced by different factors that limit the number of objects that can 
be observed for a certain period of time. Magnitude limitations imposed by the telescope size, the zenith distance of the asteroids at the time of the observations (i.e., in order to minimize distortion from 
atmospheric extinction the asteroid must be observed at rather small airmass $<$ 1.5), and the proximity of the targets to the galactic plane, are among the factors that normally reduce an initial list of several tens of 
targets to only a few of them. In Sect. 2 we describe the observations and data reduction procedure. In Sect. 3 we present the results, including the taxonomic classification and mineralogical analysis of the 
asteroids. In Sect. 4, we present and summarize our conclusions.






\section{Observations and data reduction}

\subsection{Visible spectra}

The VIS spectroscopic observations were conducted with the New Technology Telescope (NTT) at La Silla, Chile on August 2010, and with the 2.2 m telescope at Calar Alto, Spain, on October 2010. 

The VIS spectra obtained with the NTT were acquired using the EFOSC2 spectrograph, equipped with a 2048$\times$2048 pixels Loral/Lesser CCD (detector \# 40), with a pixel size of 15 $\mu$m, corresponding 
to 0".12/pixel. The disperser element used was the Grism \#1 with a 5" slit width, providing an effective wavelength range of  $\sim$ 0.43-0.95 $\mu$m. The spectrograph used with the 2.2 m telescope was 
CAFOS. Due to a technical problem two different detectors were used during the observing run at Calar Alto. The first two nights (October 05 and 06) the Lor\#11i 2k$\times$2k chip (pixel size of 15 $\mu$m) was used. 
During the last night (October 07), data were acquired with the SITe-1d 2k$\times$2k chip (24 $\mu$m pixel, i.e., 0".53/pixel). This is the standard CCD for CAFOS. The disperser element was the Grism R-400 with a 
5" slit width. The covered wavelength range is $\sim$ 0.5-0.92 $\mu$m. 

Asteroid spectra were acquired using a nodding technique in which the object is alternated between two different slit positions (A and B) following the sequence ABBA. During the observations the slit was 
oriented along the parallactic angle in order to minimize the effects of differential atmospheric refraction. To obtain the relative reflectance, solar analog stars were observed at similar air masses as the 
asteroids. For each night, flat fields, bias and arc line spectra were acquired. Observational circumstances for the studied objects are presented in Table \ref{t:Table1}.

The NTT is capable of differential tracking, however previous studies found that for exposure times longer than $\sim$ 10 minutes, NEAs tend to move out of the slit \citep{2006A&A...451..331M}. Therefore, exposures 
were limited to 600 s. In the case of the 2.2 m telescope, due to its excellent tracking capability the exposure time for the asteroids was 900 s. 

The data reduction was performed using ESO-MIDAS following the same procedure described in \citep[e.g.,][]{2006A&A...451..331M, 2010Icar..208..252N}. The steps involved in the reduction process 
are: (1) sky background removal by subtracting A from B and B from A, (2) flat-field correction, (3) median filtering for removal of cosmic hits, (4) extraction of the one-dimensional spectra, (5) wavelength calibration 
using Helium and Argon arc spectra (for the NTT data) and HgHeRb arc spectra (for the Calar Alto data), (6) extinction correction (for the NTT data), using La Silla's mean extinction curve \citep{1977Msngr..11....7T}, (7)  
co-adding of individual spectra to increase the S/N, (8) division of the asteroids spectrum by the spectrum of a solar analog star, and (9) normalization of the spectra to unity at 0.55 $\mu$m. 

\begin{figure*}[!ht]
\begin{center}
\includegraphics[height=11cm]{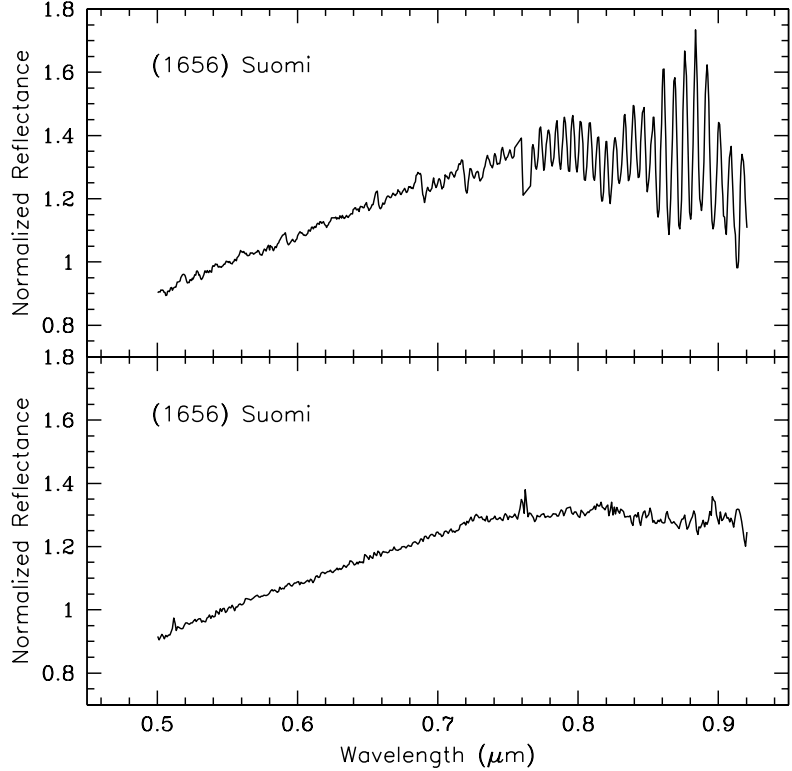}
\caption{\label{f:1656_comp} {\small The spectrum of asteroid (1656) Suomi obtained with the 2.2 m telescope using two different detectors, the Lor\#11i (top panel) and the SITe-1d 
(bottom panel). The spectrum obtained with the Lor\#11i CCD exhibits strong fringing longwards of 0.7 $\mu$m. In the present work we only report the data obtained with the SITe-1d CCD.}}
\end{center}
\end{figure*}

After reducing the data we noticed that the spectra obtained with the 2.2 m telescope and the Lor\#11i CCD exhibit strong fringing 
longwards of 0.7 $\mu$m. This can be seen in Fig. \ref{f:1656_comp} that shows the spectrum of asteroid (1656) Suomi obtained with the two detectors used during the observations, the Lor\#11i (top panel) and the 
SITe-1d (bottom panel). Since the fringing pattern could not be removed during the reduction process, in the present work we only report the data obtained with the SITe-1d CCD.

\subsection{Near-infrared spectra}

The NIR observations were conducted remotely with the NASA Infrared Telescope Facility (IRTF) on Mauna Kea, Hawai'i, on January 2007, October 2009 and July 2010. NIR spectra were obtained with the 
SpeX instrument \citep{2003PASP..115..362R}, equipped with a Raytheon Aladdin 3 1024$\times$1024 InSb array. Spectral data were acquired using Spex in its low resolution (R$\sim$150) prism mode with a 
0.8" slit width. The covered wavelength range is $\sim$ 0.7-2.5 $\mu$m. The procedure to acquire NIR spectra is essentially the same to that followed to acquire VIS  spectra. The main difference is that in addition to the 
solar analogs, and in order to correct for telluric water vapor features, local standard stars are also observed. Due to a higher and more variable background at NIR wavelengths exposures were limited to 120 s. SpeX 
data were reduced with Spextool \citep{2004PASP..116..362C}. NIR spectra are normalized to unity at 1.4 $\mu$m. Observational circumstances for the studied asteroids are presented in Table \ref{t:Table1}.

\begin{table}[!ht]
\caption{\label{t:Table1} {\small Observational circumstances. The columns in this table are: object number and designation, orbit, UT date, telescope and instrument used, number of exposures, phase angle 
($\alpha$), V-magnitude at the time of observation, air mass and solar analog used. Asteroids with the (*) symbol have a MOID $\le$ 0.05 AU and therefore are also classified as PHAs. The integration time with 
NTT/EFOSC2 was 600 s, with CA 2.2 m/CAFOS was 900 s, and with IRTF/SpeX was 120 s.}}
\begin{center}\footnotesize
\hspace*{-1.5cm}
\begin{tabular}{|c|c|c|c|c|c|c|c|c|}

\hline
Object&Orbit& UT date&Telescope and & Exp & $\alpha$ ($^\mathrm{o}$)&mag. (V)& Air mass & Solar analog   \\ 
 & & & instrument & & & & & \\ \hline
1656 Suomi&MC&07-Oct-2010&CA 2.2 m/CAFOS&4&10.4&14.5&1.52&16 Cyg B \\
2577 Litva&MC&17-Aug-2010&NTT/EFOSC2&3&12.5&15.8&1.36&Land 115-271 \\
5407 (1992 AX)&MC&12-Aug-2010&NTT/EFOSC2&4&22.8&16.6&1.06&Land 112-1333 \\
5407 (1992 AX)&MC&07-Jan-2007&IRTF/SpeX&20&16.2&15.8&1.25&Hyades 64 \\ 
22449 Ottijeff&MC&12-Aug-2010&NTT/EFOSC2&4&10.5&17.1&1.24&Land 112-1333 \\  
47035 (1998 WS)&MC&13-Aug-2010&NTT/EFOSC2&6&8.7&17.7&1.22&HD 1835 \\  
4055 Magellan&Amor&12-Aug-2010&NTT/EFOSC2&2&52.8&15.7&1.49&Land 112-1333 \\ 
4055 Magellan&Amor&21-July-2010&IRTF/SpeX&10&49.4&15.9&1.02&16 Cyg B \\  
19764 (2000 NF5)&Amor&12-Aug-2010&NTT/EFOSC2&3&36.8&16.5&1.38&Land 112-1333 \\ 
89830 (2002 CE)&Amor$^{(*)}$&07-Oct-2010&CA 2.2 m/CAFOS&2&68.1&16.3&1.41&Hyades 64 \\  
138404 (2000 HA24)&Apollo$^{(*)}$&17-Aug-2010&NTT/EFOSC2&5&2.7&16.9&1.08&Land 112-1333 \\
143381 (2003 BC21)&Amor&17-Aug-2010&NTT/EFOSC2&4&12.7&17.2&1.51&Land 115-271 \\
143381 (2003 BC21)&Amor&07-Oct-2010&CA 2.2 m/CAFOS&4&27.4&16.6&1.20&Hyades 64 \\
159609 (2002 AQ3)&Amor&16-Aug-2010&NTT/EFOSC2&4&38.8&16.7&1.25&Land 112-1333 \\
164121 (2003 YT1)&Apollo$^{(*)}$&20-Oct-2009&IRTF/SpeX&20&70.4&14.8&1.24&Hyades 64 \\ 
241662 (2000 KO44)&Amor&13-Aug-2010&NTT/EFOSC2&2&54.7&17.1&1.35&HD 1835 \\
2007 ML13&Amor&17-Aug-2010&NTT/EFOSC2&6&14.7&18.0&1.12&Land 112-1333 \\ \hline

\end{tabular}\hspace*{-1.5cm}
\end{center}
\end{table}


\section{Results}

\subsection{Taxonomic classification}

\begin{table}[!ht]
\caption{\label{t:Table2} {\small Taxonomic classification (Bus) of the observed asteroids. The measured spectral slope for each asteroid is given. If the asteroid has been classified by 
previous work the reference is indicated by (a) \citep{2002Icar..158..146B, 2002Icar..158..106B}, (b) \citep{2004Icar..172..179L} and (c) \citep{2004Icar..170..259B}. Asteroid 143381 (2003 BC21) was observed with two 
telescopes, the (*) symbol was assigned to the spectrum obtained with the 2.2 m telescope at Calar Alto. In the case of asteroid 164121 (2003 YT1) there is no VIS spectrum, therefore it is not possible to 
classify it in the Bus system. However, based on its NIR spectral characteristics we concluded that this NEA is a V-type.}}
\begin{center}
\begin{tabular}{|c|c|c|c|}
 \hline
Object&Slope ($\mu$m$^{-1}$) &Tax (previous work)&Tax (present work) \\ \hline
1656 Suomi&1.080&Ld$^{(b)}$&S-complex \\
2577 Litva&0.819&Sl$^{(b)}$&Sl \\
5407 (1992 AX)&0.575&Sk$^{(a)}$&S \\ 
22449 Ottijeff&0.567&S$^{(a)}$&S \\
47035 (1998 WS)&0.462&Sr$^{(a)}$&Sr \\ 
4055 Magellan&-0.153&V$^{(b,c)}$&V \\ 
19764 (2000 NF5)&0.441&---&S\\ 
89830 (2002 CE)&0.382&---&S \\ 
138404 (2000 HA24)&0.449&---&S \\
143381 (2003 BC21)&0.528&---&S \\
143381 (2003 BC21)$^{(*)}$&0.354&---&S \\
159609 (2002 AQ3)&0.419&---&Sk \\
164121 (2003 YT1)&---&---&V \\  
241662 (2000 KO44)&1.066&---&Sa  \\ 
2007 ML13&0.797&---&Sl \\  \hline

\end{tabular}
\end{center}
\end{table}

\begin{figure*}[!ht]
\begin{center}
\includegraphics[height=13cm]{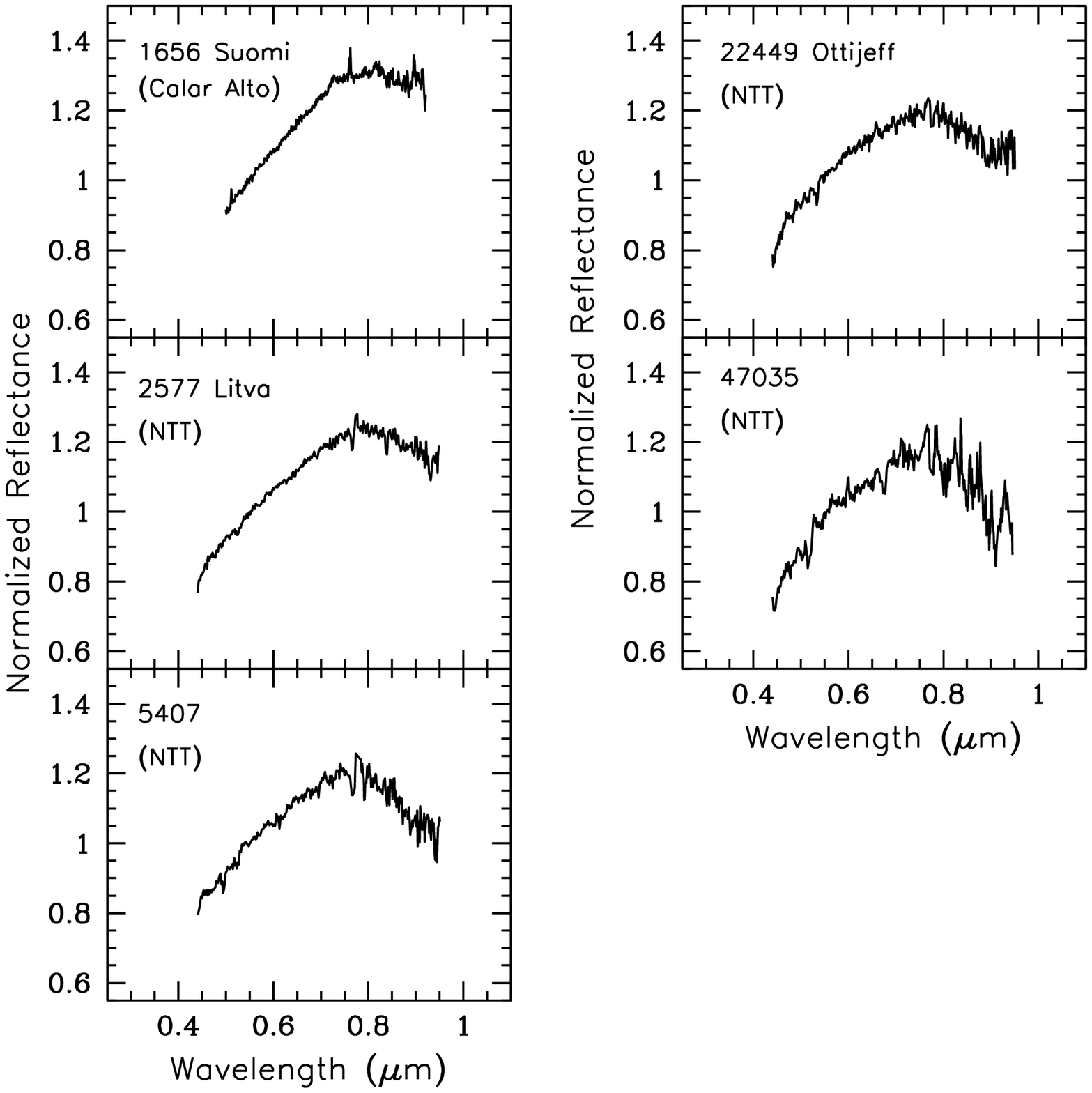}
\caption{\label{f:Allspec1} {\small Spectra of MC asteroids obtained with the NTT and the 2.2 m telescope. The numerical designation for each asteroid is given. All spectra are normalized to unity at 0.55 $\mu$m.}}
\end{center}
\end{figure*}

\begin{figure*}[!ht]
\begin{center}
\includegraphics[height=13cm]{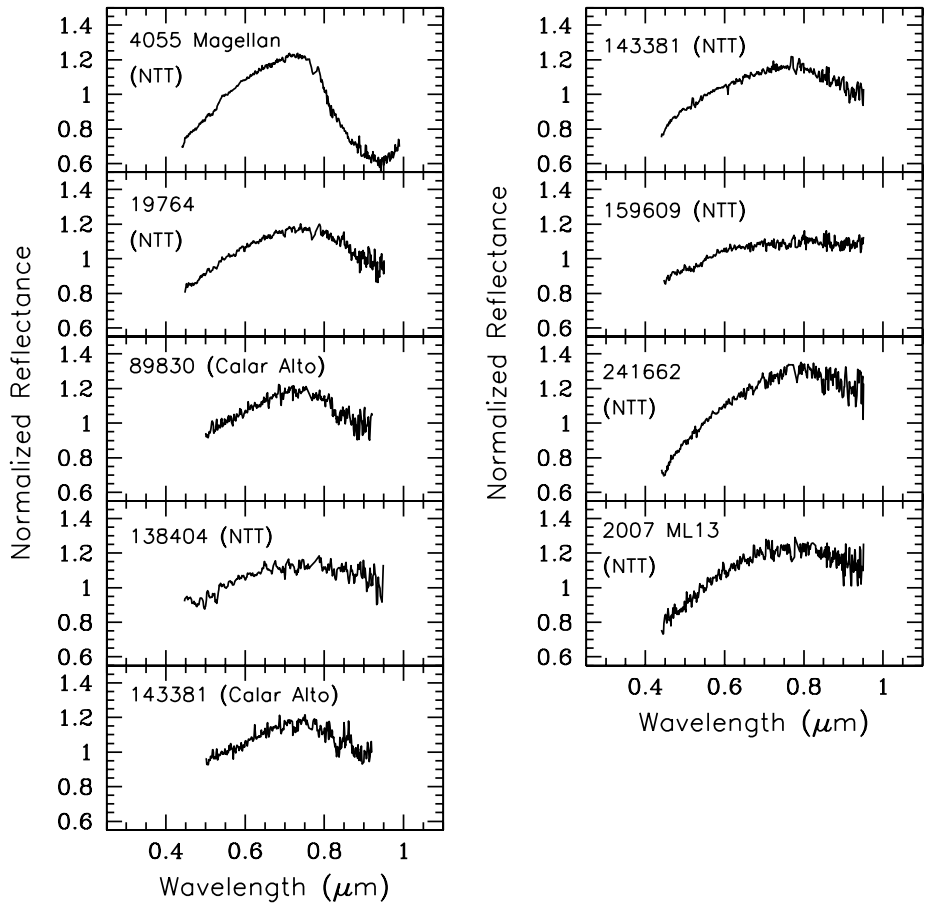}
\caption{\label{f:Allspec2} {\small Spectra of NEAs obtained with the NTT and the 2.2 m telescope. The numerical designation for each asteroid is given. All spectra are normalized to unity at 0.55 $\mu$m. Asteroid 
143381 (2003 BC21) was observed with both telescopes, due to bad weather conditions the spectrum obtained with the 2.2 m telescope at Calar Alto looks noisier than the one obtained with the NTT.}}
\end{center}
\end{figure*}

The taxonomic classification of the asteroids was made following the system developed by \citet{1999PhDT........50B}, which is comprised of 26 classes that include three major 
complexes (S-, C- and X- complex). The classification was obtained using a computer program that includes the steps described in \citet{1999PhDT........50B}. Asteroid VIS spectra are shown in 
Figs. \ref{f:Allspec1} and \ref{f:Allspec2}. The taxonomic classification of the observed asteroids is presented in Table \ref{t:Table2}. If the asteroid was previously classified the taxonomic class and reference 
are indicated. 

\begin{figure*}[!ht]
\begin{center}
\includegraphics[height=12cm]{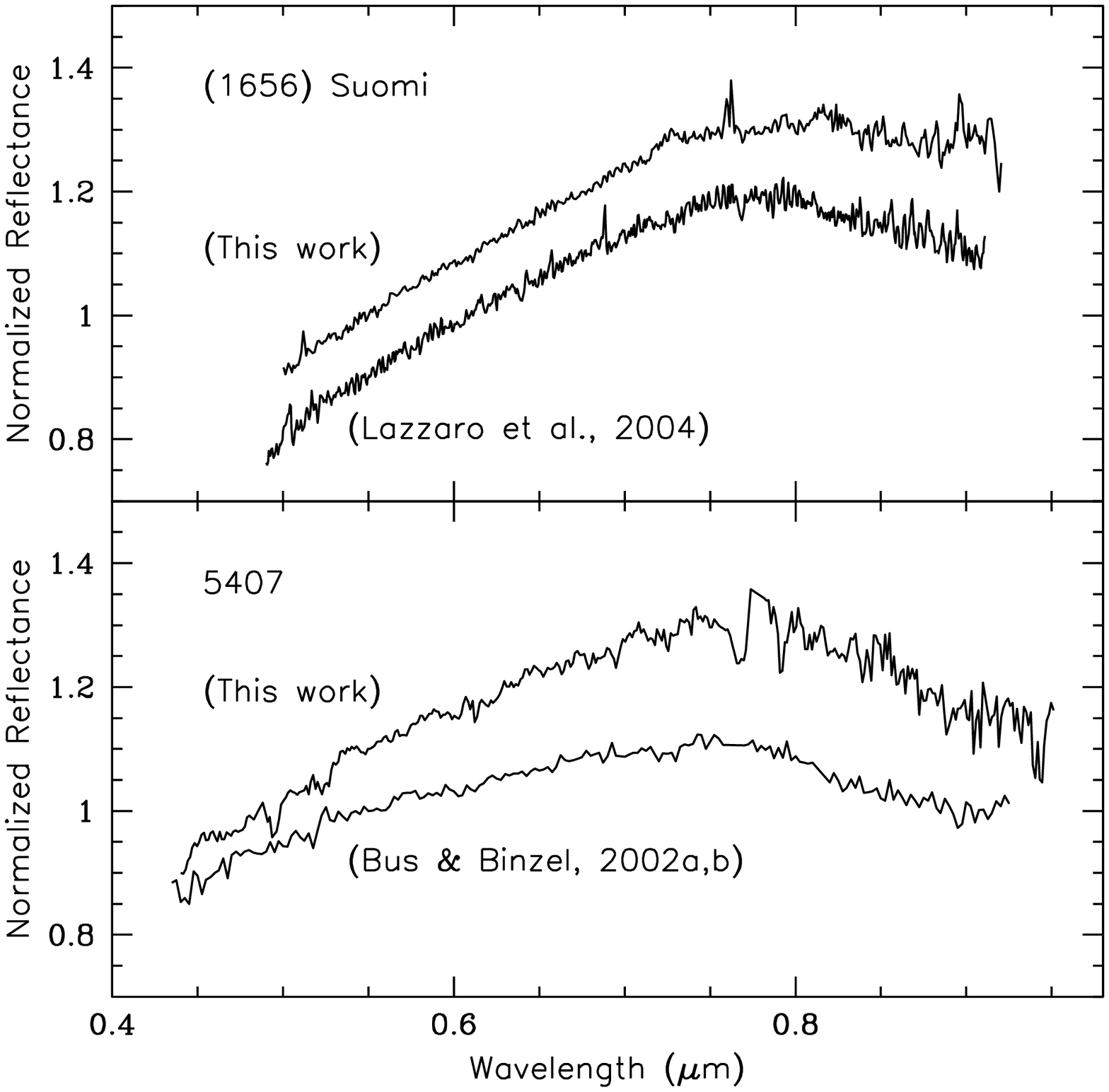}
\caption{\label{f:1656_5407} {\small Spectra of asteroid (1656) Suomi (top panel) and 5407 (bottom panel) compared with those obtained by \citet{2004Icar..172..179L} and 
\citet{2002Icar..158..146B, 2002Icar..158..106B}, respectively. Spectra are normalized to unity at 0.55 $\mu$m, and are offset for clarity.}}
\end{center}
\end{figure*}

Five of the observed objects are MC asteroids, and according to our analysis all belong to the S-complex (see Tables \ref{t:Table1} and \ref{t:Table2}). These objects have 
been observed and classified by previous studies \citep{2002Icar..158..146B, 2002Icar..158..106B,  2004Icar..172..179L,  2004Icar..170..259B}. Discrepancies between the taxonomic classification 
assigned by us and that by previous work were found for two of the asteroids (1656 Suomi and 5407). According to our results asteroid (1656) Suomi belongs to the S-complex, however due to the scattering in 
the data beyond 0.8 $\mu$m we were not able to establish its subtype within the S-complex. \citet{2004Icar..172..179L} classified this asteroid as Ld. In the 
case of asteroid 5407 it was classified by us as S-type, while \citet{2002Icar..158..146B, 2002Icar..158..106B} classified it as Sk (same complex, different subtype). Figure \ref{f:1656_5407} shows the spectra of 
asteroid (1656) Suomi and 5407 compared with those obtained by \citet{2004Icar..172..179L} and \citet{2002Icar..158..146B, 2002Icar..158..106B}, respectively. As noted by \citet{2002aste.conf..169B}, these 
kind of discrepancies are often reported when the spectra of asteroids observed by different surveys are compared. These differences could be attributed to several factors like instrumental 
effects and viewing geometry effects \citep{2002aste.conf..169B}. The latter can manifest themselves as phase reddening, which produces an increase of the spectral slope and variations in the intensity of 
the absorption bands \citep{2012Icar..220..36S}. The poor weather conditions during our observing run at Calar Alto could also contribute to explain the difference between the taxonomic classification 
assigned by us and that by previous studies.

Of the nine NEAs observed, only two, 4055 Magellan and 164121 (2003 YT1), don't belong to the S-complex. The VIS spectrum of 4055 Magellan exhibits a very deep 1-$\mu$m absorption band 
characteristic of pyroxene assemblages  \citep[e.g.,][]{1991JGR....9622809C, 2002aste.conf..183G, 2010Icar..208..773M}. Our classification of this asteroid as a V-type is consistent with previous 
studies \citep{2004Icar..172..179L, 2004Icar..170..259B}. For 164121 (2003 YT1) we only have a NIR spectrum, therefore we couldn't classify it under the Bus system. However, based on 
its mineralogical analysis (see below) we found that this NEA is a V-type asteroid as well (see Figure \ref{f:VNIRspectra7}). NEAs 19764 (2000 NF5), 138404 (2000 HA24), 143381 (2003 BC21) and 89830 (2002 CE) 
were classified as S-types, while 241662 (2000 KO44), 159609 (2002 AQ3) and 2007 ML13 were classified as Sa, Sk and Sl, respectively.    

\clearpage

\subsection{Mineralogical analysis}

Our observations with the IRTF include NIR spectra of asteroids 4055 Magellan, 164121 (2003 YT1), and 5407 (1992 AX). For three of the NEAs that we obtained visible spectra, 241662 (2000 KO44), 
19764 (2000 NF5) and 138404 (2000 HA24), there is also available NIR spectra (0.8-2.5 $\mu$m) from the MIT-UH-IRTF Joint Campaign for NEO Spectral Reconnaissance (NEOSR). These spectra were acquired 
with the IRTF and SpeX. All the data (with the exception of 164121 (2003 YT1)) were combined with the VIS spectra to increase the wavelength coverage, allowing the mineralogical characterization of these 
asteroids. For that purpose, spectral band parameters, band centers and Band Area Ratios (BAR) for each VIS-NIR spectrum were measured in the same way as in \citet{2012Icar..220..36S}. In the case of 
NEA 164121 (2003 YT1), the wavelength range covered ($\sim$ 0.7-2.5 $\mu$m) was sufficient to allow us to measure these band parameters. 

\begin{figure*}[!ht]
\begin{center}
\includegraphics[height=12cm]{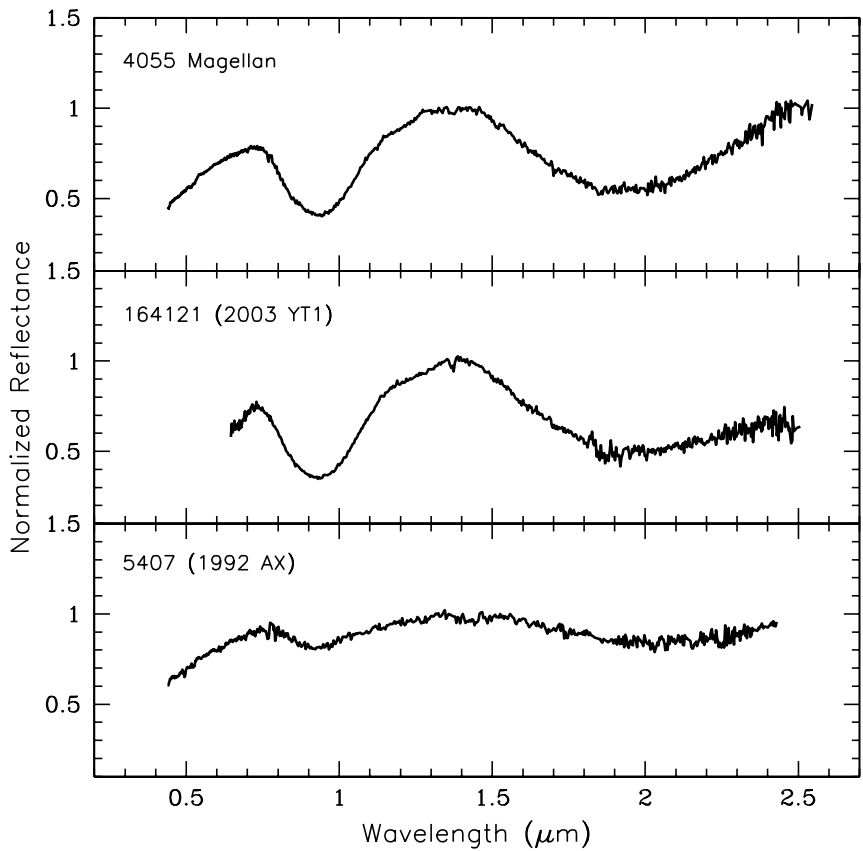}
\caption{\label{f:VNIRspectra7} {\small VIS-NIR spectra of asteroids 4055 Magellan (top panel) and 5407 (1992 AX) (bottom panel), and NIR spectrum of 164121 
(2003 YT1) (middle panel). All spectra are normalized to unity at 1.4 $\mu$m.}}
\end{center}
\end{figure*}

\begin{figure*}[!ht]
\begin{center}
\includegraphics[height=12cm]{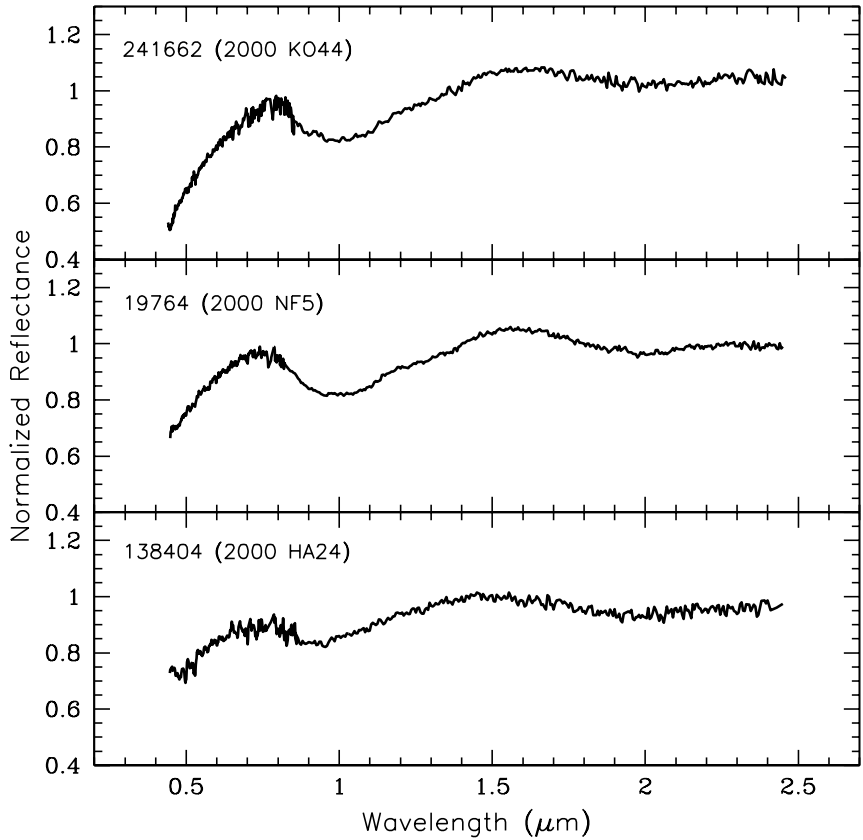}
\caption{\label{f:VNIRspectra8} {\small VIS-NIR spectra of asteroids 241662 (2000 KO44) (top panel), 19764 (2000 NF5) (middle panel) and 138404 (2000 HA24) 
(bottom panel). NIR asteroid spectra were obtained from the NEOSR survey (http://smass.mit.edu/minus.html). All spectra are normalized to unity at 1.4 $\mu$m.}}
\end{center}
\end{figure*}

The VIS-NIR spectra of asteroids 4055 Magellan and 5407 (1992 AX), and the NIR spectrum of 164121 (2003 YT1) are shown in Fig. \ref{f:VNIRspectra7}. The spectra of these objects show 
absorption bands characteristics of pyroxene assemblages, one centered near 0.9-1 $\mu$m (Band I) and the other at $\sim$ 1.9-2 $\mu$m (Band II). These absorption bands are caused by the presence of 
Fe$^{2+}$ cations located in the M2 crystallographic site \citep{1974JGR....79.4829A, 1975Adams, 1993macf.book.....B}. The VIS-NIR spectra of asteroids 241662 (2000 KO44), 19764 (2000 NF5) and 138404 (2000 
HA24) (shown in Fig. \ref{f:VNIRspectra8}) exhibit the two absorption bands characteristics of olivine-orthopyroxene assemblages. 

As stated earlier, the VIS-NIR spectra of 4055 Magellan and 5407 (1992 AX), and the NIR spectrum of 164121 (2003 YT1) show spectral characteristics of basaltic asteroids. Therefore, to the band parameters, we 
applied the temperature corrections derived by \citet{2012Icar..217..153R}. In addition, the pyroxene compositions of these objects were determined using the equations derived by \citet{2007LPI....38.2117B}. With 
these equations we can determine the molar contents of ferrosilite (Fs) and wollastonite (Wo) for basaltic asteroids. 

For asteroids 241662 (2000 KO44), 19764 (2000 NF5) and 138404 (2000 HA24), which exhibit spectral characteristics of olivine-orthopyroxene assemblages, temperature corrections to band parameters were applied 
using the equations obtained by \citet{2012Icar..220..36S}. The mafic silicate compositions of these objects were determined using the equations derived by \citet{2010Icar..208..789D}. With these equations it is 
possible to determine the molar contents of fayalite (Fa) in olivine and ferrosilite (Fs) in pyroxene in olivine-orthopyroxene assemblages. In addition, the mineral abundances $(ol/(ol + px))$ were calculated from the 
BAR values using the relationship found by \citet{2010Icar..208..789D}. This relationship was obtained from the analysis of ordinary chondrites and therefore can be applied to asteroids with similar mineralogies. 
Spectral band parameters and mineral chemistry values for the studied asteroids are presented in Table \ref{t:Table3}.

\begin{figure*}[!ht]
\begin{center}
\includegraphics[height=12cm]{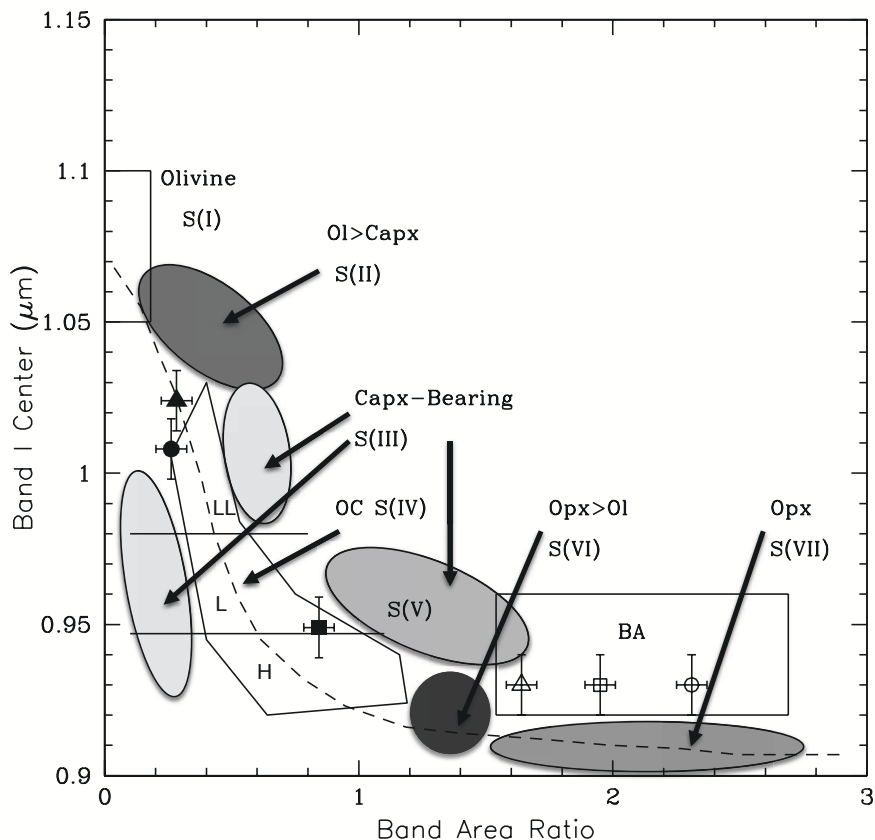}
\caption{\label{f:BI_BARH2} {\small Plot of the Band I center versus BAR for 241662 (2000 KO44) (filled triangle), 19764 (2000 NF5) (filled circle), 138404 (2000 HA24) (filled square), 
4055 Magellan (open square), 5407 (1992 AX) (open circle), and 164121 (2003 YT1) (open triangle). The average 1-$\sigma$ error bars were calculated as in \citet{2012Icar..220..36S}. The rectangular zone (BA) 
includes the pyroxene-dominated basaltic achondrite assemblages \citep{1993Icar..106..573G}. The polygonal region, corresponding to the S(IV) subgroup, represents the mafic silicate components of ordinary 
chondrites (OC). The dashed curve indicates the location of the olivine-orthopyroxene mixing line \citep{1986JGR....9111641C}. The horizontal lines represent the approximate boundaries for ordinary chondrites found 
by \citet{2010Icar..208..789D}.}}
\end{center}
\end{figure*}

Apart from the mineralogical analysis, the spectral band parameters (Band I center and BAR) were also used to classify these asteroids in the system introduced by \citet{1993Icar..106..573G}, 
which divides the S-population into seven main compositional subgroups designated S(I)-S(VII). Fig. \ref{f:BI_BARH2} shows the measured Band I center versus BAR for 4055 Magellan (open square), 5407 
(1992 AX) (open circle) and 164121 (2003 YT1) (open triangle). These three asteroids are located in the rectangular zone (BA), which includes the pyroxene-dominated basaltic achondrite 
assemblages \citep{1993Icar..106..573G}. Within the uncertainties, asteroids 241662 (2000 KO44), 19764 (2000 NF5) and 138404 (2000 HA24) are classified as S(IV) under the system 
of  \citet{1993Icar..106..573G}. In Fig. \ref{f:BI_BARH2} are also depicted the measured Band I center versus BAR for 241662 (2000 KO44) (filled triangle), 19764 (2000 NF5) (filled circle) and 138404 (2000 HA24) 
(filled square).
    
\clearpage

\begin{figure*}[!ht]
\begin{center}
\includegraphics[height=12cm]{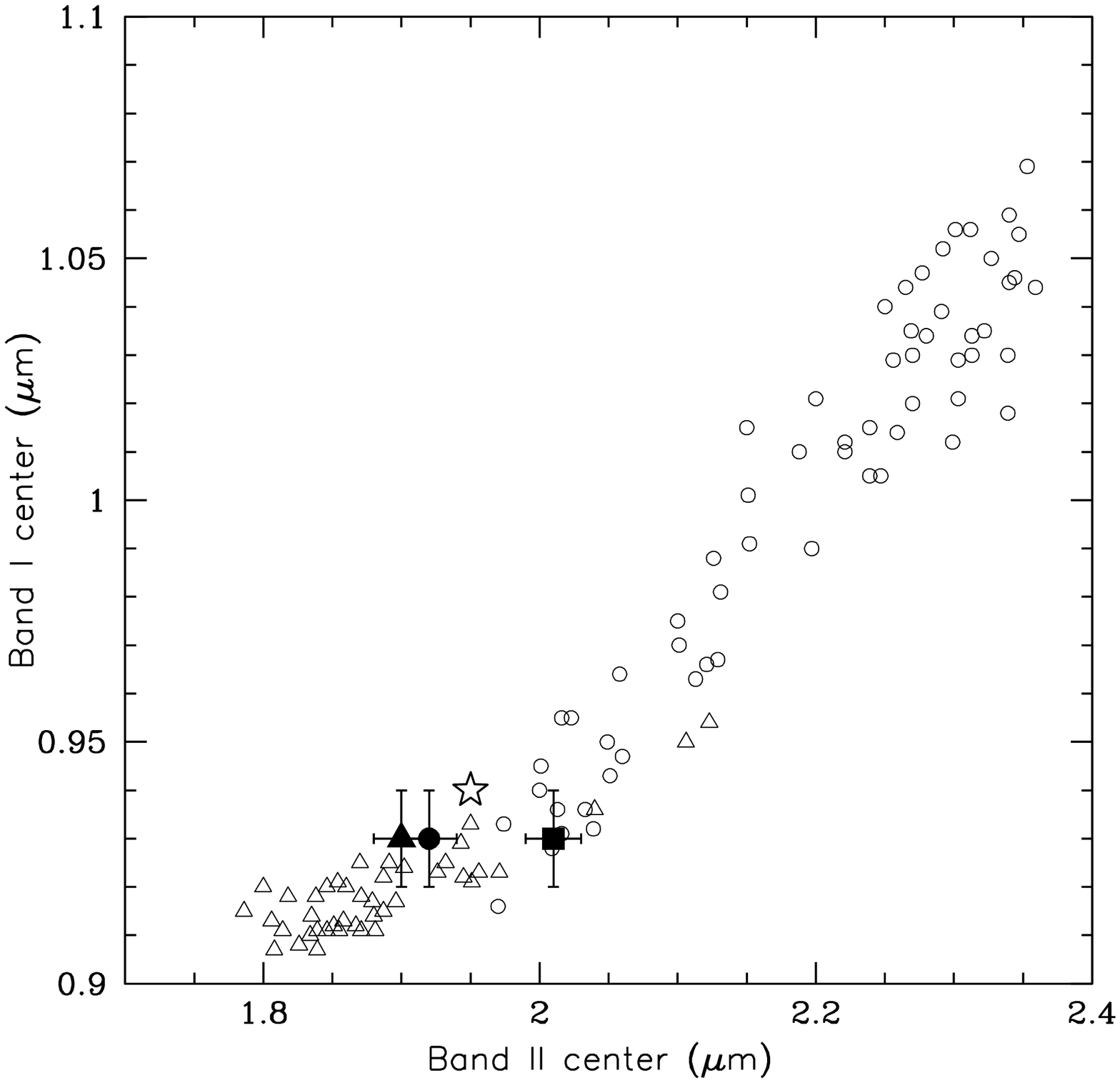}
\caption{\label{f:BI_BII2} {\small Plot of Band I center vs. Band II center for 4055 Magellan (filled circle), 164121 (2003 YT1) (filled triangle), and 5407 (1992 AX) (filled square). The average 1-$\sigma$ error bars 
were calculated as in \citet{2012Icar..220..36S}. Open triangles represent measured band centers for orthopyroxenes from \citet{1974JGR....79.4829A}, open circles correspond to measured band centers for 
clinopyroxenes from \citet{1991JGR....9622809C}. For comparison we plotted the band centers of (4) Vesta (star symbol) measured by \citet{2012Icar..217..153R}. The error bars are smaller than the symbol.}}    
\end{center}
\end{figure*}
\begin{figure*}[!ht]
\begin{center}
\includegraphics[height=12cm]{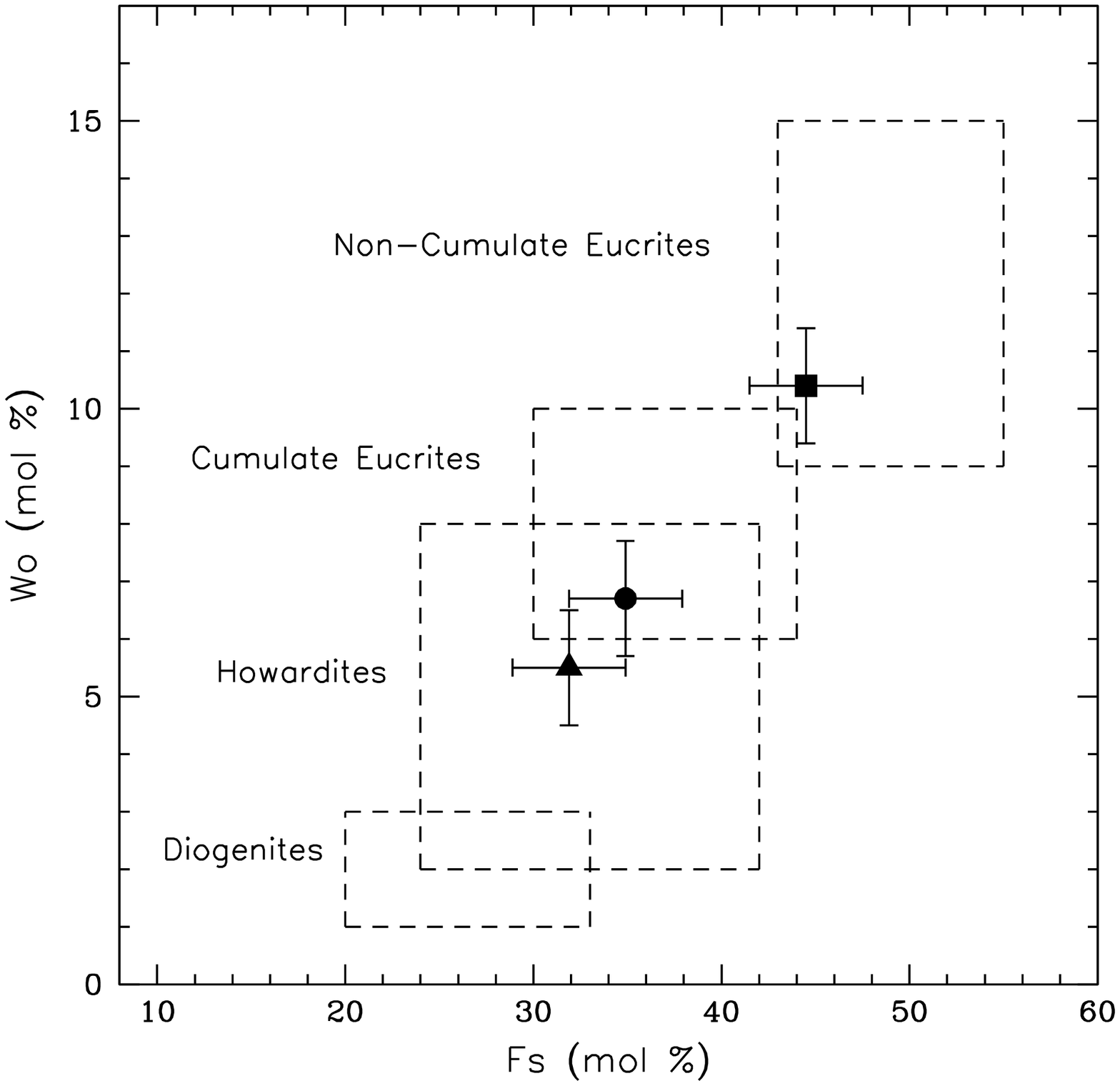}
\caption{\label{f:Wo_Fs2} {\small Molar contents of Wo vs. Fs for NEAs 4055 Magellan (filled circle), 164121 (2003 YT1) (filled triangle), and 5407 (1992 AX) (filled square). The error bars correspond to the values 
determined by \citet{2007LPI....38.2117B}, 3 mol \% for Fs and 1 mol \% for Wo. The approximated range of pyroxene chemistries for howardites, non-cumulate eucrites, cumulate eucrites, and diogenites from \citet{1998PlanetaryMaterialsM} are indicated as dashed-line boxes.}}    
\end{center}
\end{figure*}

For NEA 4055 Magellan, the Band I center is located at 0.93$\pm0.01$ $\mu$m and the Band II center is at 1.92$\pm$0.02 $\mu$m with a BAR of 1.95$\pm$0.06. After plotting the band centers of this asteroid on 
the Band I vs. Band II center plot (Fig. \ref{f:BI_BII2}, filled circle), we found that this object has a surface assemblage dominated by orthopyroxene. For comparison we also plotted the band centers of asteroid (4) 
Vesta (star symbol) measured by \citet{2012Icar..217..153R}. The spectrum of this asteroid shows an inflection at $\sim$ 1.3 $\mu$m, which is visible in Type A clinopyroxene, olivine-orthopyroxene mixtures and 
plagioclase \citep{1993macf.book.....B}. From our calculation we determined that the pyroxene chemistry of 4055 Magellan is Fs$_ {35}$En$_ {58}$Wo$_ {7}$. These values are consistent with 
those measured for howardites (Fs$_ {24-42}$,Wo$_ {2-8}$), and cumulate eucrites (Fs$_ {30-44}$,Wo$_ {6-10}$)  \citep{1998PlanetaryMaterialsM}. Our results are in agreement with those found 
by \citet{2009M&PS...44.1331B}, which gave a pyroxene chemistry of Fs$_ {36}$En$_ {57}$Wo$_ {7}$ for Magellan. In Fig. \ref{f:Wo_Fs2} we plot the molar content of Wo versus Fs for 4055 Magellan, depicted as 
a filled circle. The approximated range of pyroxene chemistries found for howardites, eucrites and diogenites are shown as dashed-line boxes.


For the PHA 164121 (2003 YT1), the Band I and Band II centers are located at 0.93$\pm$0.01 $\mu$m and 1.90$\pm$0.02 $\mu$m, respectively. The calculated BAR is 1.64$\pm$0.06. Similarly to 4055 Magellan, 
this asteroid exhibits a surface assemblage dominated by orthopyroxene (filled triangle in Fig. \ref{f:BI_BII2}). Like Magellan, the spectrum of 164121 (2003 YT1) shows an inflection at $\sim$ 1.3 $\mu$m. For this 
NEA the pyroxene chemistry was determined to be Fs$_ {32}$En$_ {62}$Wo$_ {6}$. These values are consistent with the measured ranges for both howardites and cumulate eucrites, and are similar to the values reported 
by \citet{2004DPS....36.2809A}, who estimated a pyroxene chemistry of Fs$_ {32}$En$_ {60}$Wo$_ {8}$ for this asteroid. The molar contents of Wo and Fs for 164121 (2003 YT1) are shown in 
Fig. \ref{f:Wo_Fs2} (filled triangle).  


The measured Band I and Band II centers of the MC 5407 (1992 AX) are located at 0.93$\pm$0.01 $\mu$m and 2.01$\pm$0.02 $\mu$m, respectively, with a BAR of 2.31$\pm$0.06. The position of the band centers, 
which are represented in Fig. \ref{f:BI_BII2} as a filled square, indicates that this asteroid has a surface assemblage possibly dominated by clinopyroxenes. As in the previous cases, the pyroxene chemistry of the 
asteroid was calculated, given the value of Fs$_ {45}$En$_ {45}$Wo$_ {10}$. This pyroxene chemistry is consistent with the range estimated for non-cumulate eucrites 
(Fs$_ {43-55}$, Wo$_ {9-15}$)  \citep{1998PlanetaryMaterialsM}. The molar contents of Wo and Fs calculated for 5407 (1992 AX) are indicated in Fig. \ref{f:Wo_Fs2} with a filled square. The surface composition of this 
asteroid contrast with its taxonomic classification derived from the VIS spectrum, which classify it as a S-type. However, the analysis of NIR spectra of individual S-type asteroids has proved that this class can include both 
differentiated and primitive objects \citep{1993Icar..106..573G}. Asteroid 5407 represents one of those cases in which a S-type exhibits signs of differentiation. This object is also a good example of the limitations of 
taxonomic classification, and shows how the assumption that objects belonging to the same class share similar mineralogies can lead to a misinterpretation about the real nature of the asteroid. 


\begin{figure*}[!ht]
\begin{center}
\includegraphics[height=12cm]{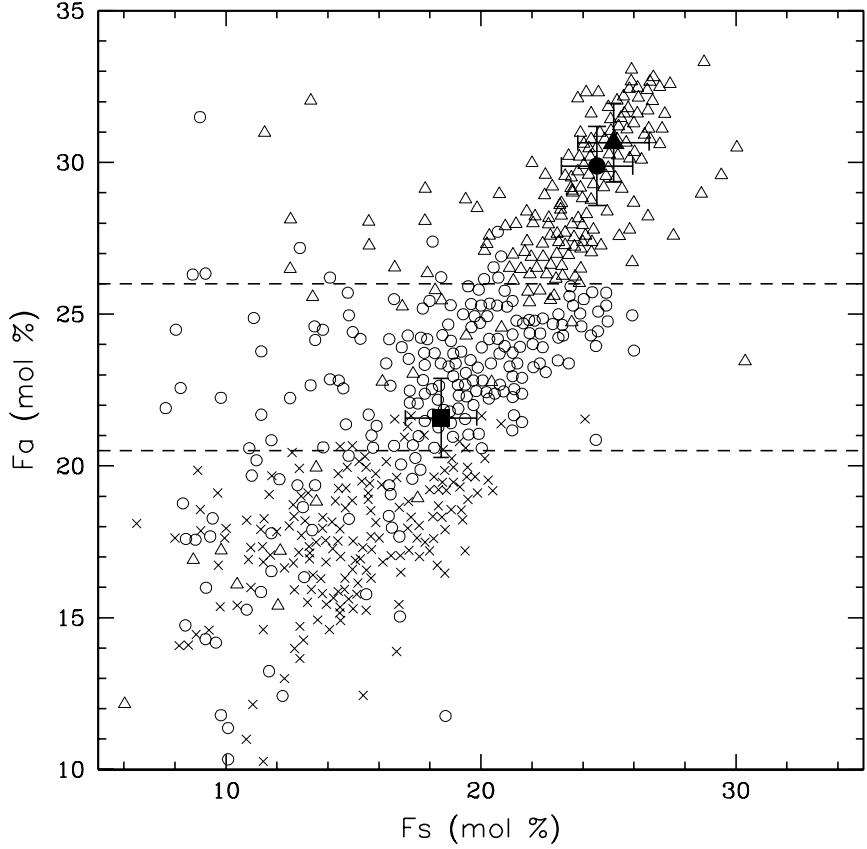}
\caption{\label{f:Fa_Fs2} {\small Molar contents of Fa vs. Fs for NEAs 241662 (2000 KO44) (filled triangle), 19764 (2000 NF5) (filled circle), 138404 (2000 HA24) (filled square), along 
with the values for LL (open triangles), L (open circles) and H (x) ordinary chondrites from \citet{2011Sci...333.1113N}. The error bars correspond to the values determined by 
\citet{2010Icar..208..789D}, 1.4 mol \% for Fs and 1.3 mol \% for Fa. The horizontal dashed-lines represent the approximate boundaries for LL, L, and H ordinary chondrites.}}    
\end{center}
\end{figure*}

For NEA 241662 (2000 KO44), the Band I center is located at 1.02$\pm$ 0.01 $\mu$m and the Band II center is at 1.99$\pm$ 0.02 $\mu$m with a BAR of 0.28$\pm$0.06. The Band I center and BAR values of this 
object place it in the upper limit of the polygonal region defined by \citet{1993Icar..106..573G}, and just on the olivine-orthopyroxene mixing line of \citet{1986JGR....9111641C} (Fig. \ref{f:BI_BARH2}). Within the error 
bars this asteroid is classified as a S(IV) in the system of \citet{1993Icar..106..573G}. Using the equations derived by \citet{2010Icar..208..789D} we determined that the olivine and pyroxene chemistries are 
Fa$_ {31}$(Fo$_ {69}$) and Fs$_ {25}$, respectively. These values are consistent with those derived for LL ordinary chondrites by \citet{2010Icar..208..789D}, which give a range of Fa$_ {(25-33)}$ for fayalite and 
Fs$_ {(21-27)}$ for ferrosilite. The molar contents of Fa vs. Fs for 241662 (2000 KO44) are shown in Fig. \ref{f:Fa_Fs2} as a filled triangle. For comparison we also plotted the values for LL (open triangles), L (open 
circles) and H (x symbols) ordinary chondrites from \citet{2011Sci...333.1113N}. From the BAR value we estimated that the olivine and pyroxene percentages for this asteroid are 0.66 and 0.34, respectively. These 
values are also in agreement with X-ray diffraction (XRD) measurements reported by \citet{2010Icar..208..789D} for LL ordinary chondrites, that give $ol/(ol+px)$ values in the range of 0.58-0.69.


The Band I center of NEA 19764 (2000 NF5) is located at 1.01$\pm$0.01 $\mu$m and the Band II center is at 1.99$\pm$0.02 $\mu$m with a BAR of 0.26$\pm$0.06. The Band II center of this asteroid is consistent with 
those of low calcium ($\sim$1.8-2.1 $\mu$m) and high calcium pyroxenes ($\sim$1.97-2.35 $\mu$m). The band parameters of 19764 (2000 NF5) are slightly offset to the left of the olivine-orthopyroxene mixing line 
(see Fig. \ref{f:BI_BARH2}). This shift could be due to the presence of small amounts of high calcium pyroxene along with olivine and orthopyroxene in the assemblage. Within the error bars, 19764 (2000 NF5) belongs 
to the S(IV) compositional subgroup. For this asteroid the calculated Fa and Fs content (Fa$_ {30}$ and Fs$_ {25}$), as well as the $ol/(ol+px)$ ratio of 0.66, fall into the range of LL ordinary chondrites. 
Fig. \ref{f:Fa_Fs2} shows the molar contents of Fa and Fs for 19764 (2000 NF5) (filled circle). 


In the case of the PHA 138404 (2000 HA24), the Band I and Band II centers are located at 0.95$\pm$0.01 $\mu$m  and 1.98$\pm$0.02 $\mu$m, respectively, with a 
BAR of 0.84$\pm$0.06. This asteroid plots in the S(IV) region (Fig. \ref{f:BI_BARH2}) at the boundary between H and L chondrites. As can be seen in Fig. \ref{f:BI_BARH2} the band parameters of this object are offset 
to the right of the olivine-orthopyroxene mixing line, which is an indication for the presence of high calcium pyroxene on the surface of the asteroid \citep{1993Icar..106..573G}. According to our calculations, the olivine 
and pyroxene chemistries for this asteroid are Fa$_ {22}$(Fo$_ {78}$) and Fs$_ {18}$, respectively. These values are in the borderline of those derived by \citet{2010Icar..208..789D} for L (Fa$_ {21-27}$ and 
Fs$_ {17-23}$) and H (Fa$_ {15-21}$ and Fs$_ {13-19}$) ordinary chondrites. The $ol/(ol+px)$ ratio of 0.52 calculated for 138404 (2000 HA24) falls into the range of H ordinary chondrites (0.46-0.60) measured 
by \citet{2010Icar..208..789D}. The derived olivine and pyroxene chemistries of 138404 (2000 HA24) are shown in Fig. \ref{f:Fa_Fs2} as a filled square.

\begin{table}[!ht]
\caption{\label{t:Table3} {\small Spectral band parameters for the asteroids studied. Molar contents of Fayalite (Fa), Ferrosilite (Fs), wollastonite (Wo), and ol/(ol+px) ratio for each asteroid (where applicable) 
are presented.}}
\begin{center}\footnotesize
\hspace*{-1cm}
\begin{tabular}{|c|c|c|c|c|c|c|c|}

\hline
Object&Band I center&Band II center&BAR&Fa&Fs&Wo&ol/(ol+px)   \\ 
 &($\mu$m)&($\mu$m)& &(mol \%)&(mol \%)&(mol \%)& \\ \hline
5407 (1992 AX)&0.93$\pm$0.01&2.01$\pm$0.02&2.31$\pm$0.06&---&44.5$\pm$3.3\%&10.4$\pm$1.1\%&--- \\ 
4055 Magellan&0.93$\pm$0.01&1.92$\pm$0.02&1.95$\pm$0.06&---&34.9$\pm$3.3\%&6.7$\pm$1.1\%&--- \\ 
19764 (2000 NF5)&1.01$\pm$0.01&1.99$\pm$0.02&0.26$\pm$0.06&29.9$\pm$1.3\%&24.6$\pm$1.4\%&---&0.66$\pm$0.03 \\ 
138404 (2000 HA24)&0.95$\pm$0.01&1.98$\pm$0.02&0.84$\pm$0.06&21.6$\pm$1.3\%&18.4$\pm$1.4\%&---&0.52$\pm$0.03 \\ 
164121 (2003 YT1)&0.93$\pm$0.01&1.90$\pm$0.02&1.64$\pm$0.06&---&31.9$\pm$3.3\%&5.5$\pm$1.1\%&--- \\ 
241662 (2000 KO44)&1.02$\pm$0.01&1.99$\pm$0.02&0.28$\pm$0.06&30.7$\pm$1.3\%&25.2$\pm$1.4\%&---&0.66$\pm$0.03 \\ \hline

\end{tabular}\hspace*{-1cm}
\end{center}
\end{table}


\clearpage

\section{Conclusions}

We have found that all observed MC asteroids, five in total, belong to the S-complex, including the S, Sr and Sl classes. The classifications assigned by us are 
consistent with previous work with the exception of two cases. This discrepancy could be attributed to atmospheric conditions or phase angle effects. Among the NEAs 
observed, seven belong to the S-complex, including the S, Sa, Sk and Sl classes. All of them have been classified for the first time in the present work. Two NEAs were classified as V-type asteroids. 

The mineralogical analysis of NEAs 4055 Magellan and 164121 (2003 YT1), showed that their pyroxene chemistries are consistent with both, howardites and cumulate eucrites. 
Howardi-tes are polymict breccia composed by a mixture of eucrite and diogenite fragments. Cumulate eucrites, on the other hand, are achondritic meteorites, which are thought to have formed in shallow 
magma chambers within the crust of their parent body, where they slowly crystallized. Howardites and eucrites along with diogenites (HED meteorites) have been linked to asteroid (4) Vesta, on the basis of their 
spectral and mineralogical similarities derived from ground-based observations \citep[e.g.,][]{1970Sci...168.1445M, 1977GeCoA..41.1271C, 1997Icar..127..130G, 2002aste.conf..573K}. This idea has been supported 
by recent spacecraft spectroscopic observations obtained by Dawn \citep[e.g.,][]{2012Sci...336..700R, 2012Sci...336..697D}, that shows Vesta as the most likely parent body of these meteorites. The Vesta asteroid 
family extends from the $\nu_{6}$ secular resonance to the 3:1 mean motion resonance with Jupiter, providing a viable route for the delivery of these objects to the near-Earth space 
\citep{1985Natur.315..731W, 2002aste.conf..395B}. Our results suggest that 4055 Magellan and 164121 (2003 YT1) are formed by regolith breccia originated in a larger asteroid that experienced extensive 
igneous processing. The pyroxene chemistries derived for these asteroids are similar to those found by \citet{2009M&PS...44.1331B} for a group of V-type NEAs. Although we cannot rule out a different parent body for 
these NEAs, these results are consistent with the aforementioned scenario, which would link Vesta with these asteroids via secular or mean motion resonances.

The mineralogical analysis of MC 5407 (1992 AX) revealed a pyroxene chemistry consistent with the range estimated for non-cumulate eucrites. These results suggest that the parent body of this asteroid 
experienced temperatures high enough to cause at least partial melting, making basaltic achondrites the most plausible meteorite analogs \citep{1993Icar..106..573G}.  

For NEAs, 241662 (2000 KO44), 19764 (2000 NF5) and 138404 (2000 HA24), the mineralogical analysis revealed ordinary chondrite-like compositions. In particular, 241662 (2000 KO44) and 
19764 (2000 NF5) showed mineral compositions consistent with olivine-rich LL ordinary chondrites. Within the error bars the olivine and pyroxene chemistries of these two asteroids, as well as the 
olivine abundances, are consistent with the values measured from the returned samples of asteroid 25143 Itokawa. The mean compositions of olivine and low-Ca pyroxene for these samples gave values of 
Fa$_ {29}$ and Fs$_ {24}$, respectively \citep{2011Sci...333.1113N}, with an olivine abundance of 64\% \citep{2011Sci...333.1125T}. The analysis of Itokawa samples also revealed that this asteroid is a 
breccia, consisting of poorly equilibrated LL4 and highly equilibrated LL5 and LL6 materials \citep{2011Sci...333.1113N}. This degree of thermal metamorphism indicates that the parent body where Itokawa originated 
must have had a diameter larger than 20 km, necessary to reach temperatures of $\sim$ 800$^{o}$C that can explain the range of petrologic types observed among the Itokawa 
particles \citep{2011Sci...333.1113N}. LL ordinary chondrites have been linked to the Flora family \citep{2008Natur.454..858V}, which is located in the 
inner part of the main belt near the $\nu_{6}$ secular resonance. Although the derived silicate compositions of 241662 (2000 KO44) and 19764 (2000 NF5) are consistent with that of 
LL ordinary chondrites, we noticed that their olivine abundances (66\%) are lower than the ranges found by \citet{2008Natur.454..858V} for LL chondrites  (70-85\%), and 
members of the Flora family (76-82\%). However, this discrepancy could be explained by the fact that \citet{2008Natur.454..858V} used a radiative transfer model 
\citep{1999Icar..137..235S} to calculate the $ol/(ol+px)$ ratios, while we used empirical spectral calibrations derived from the analysis of ordinary chondrites.

NEA 138404 (2000 HA24) is considered as a possible target for robotic and manned missions due to its low $\Delta$V \citep{2011AASL}. This asteroid showed a composition similar to L and H ordinary chondrites. If this 
NEA formed in a L ordinary chondrite parent body, then its origin could be related to the formation of the Gefion family. \citet{2009Icar..200..698N} found a possible link between this family (located at 2.7-2.82 AU) and L 
ordinary chondrites. Asteroids in the Gefion family are thought to have originated after an impact event that disrupted its parent body $\sim$ 470 million years ago \citep{1996Icar..119..182H, 2007M&PS...42..113K}. This 
group of objects exhibit spectral characteristics compatible with S-complex asteroids \citep{2002aste.conf..633C}.  According to \citet{2009Icar..200..698N} the location of the Gefion family, near the 5:2 mean motion 
resonance with Jupiter, provides a viable source for the delivery of this material to Earth. Thus, objects in the Gefion family could eventually escape from the main belt and evolve to Earth-crossing orbits.  

Another possible scenario for the origin of 138404 (2000 HA24) is that this object is a fragment of a H ordinary chondrite parent body. The main belt asteroid 6 Hebe, which is located at 2.426 AU between the 
$\nu_{6}$ secular resonance and the 3:1 mean motion resonance, shows spectral characteristics and mafic silicate compositions consistent with those of H chondrites 
\citep{1996M&PSA..31Q..47G, 1998M&PS...33.1281G}, making of this asteroid its probably parent body. Cosmic ray exposure ages (CRE) of H chondrites indicate that their parent body, or a fragment of it, experienced 
multiple collisional events, occurred $\sim$ 60, $\sim$ 33 and $\sim$ 8 million years ago \citep{1995JGR...10021247G}. Numerical experiments have shown that fragments ejected from Hebe at velocities of a few 
hundreds m/s can be transported into the resonance zones and from there into Earth-crossing orbits within a time span of $\sim$ 1 million years \citep{1993CeMDA..56..287F}. Thus, a possible scenario for the 
origin of 138404 (2000 HA24) is that this object was ejected during one of the collisional events and later transported to the near-Earth space. If 138404 (2000 HA24) is indeed a fragment of Hebe, then it could be 
also linked to some of the H ordinary chondrites found on Earth. It is believed that the latest collisional event occurred $\sim$ 8 million years ago is the source of $\sim$ 45\% of H chondrites that reach the 
Earth \citep{1995JGR...10021247G}. Since this collisional event took place within the typical lifetimes ($10^{6}-10^{7}$ years) associated with NEAs \citep{2002aste.conf..409M}, it is possible that some H chondrite 
meteorites were ejected from the surface of 138404 (2000 HA24). 



The compositional information derived in this work from the analysis of VIS-NIR spectral data, along with dynamical considerations allowed us to discuss possible scenarios for the origin of the studied asteroids, as 
well as their connection with meteorites found on Earth. Since our sample included PHAs an a possible target for robotic and manned missions, the information gathered here could eventually help in planning future 
missions, and contribute in the development of mitigation strategies.




\clearpage


{\bf{Acknowledgements}}

\

This paper is based on data obtained with the New Technology Telescope at La Silla, Chile, observations collected at the Centro Astron\'{o}mico Hispano Alem\'{a}n (CAHA), operated jointly by the Max-Planck 
Institut f$\ddot{u}$r Astronomie and the Instituto de Astrofisica de Andalucia (CSIC), and observations acquired with the Infrared Telescope Facility on Mauna Kea, Hawai'i. Part of the data used in this work were obtained 
from the MIT-UH-IRTF Joint Campaign for NEO Reconnaissance. The authors would like to thank Olivier Hainaut for his help during the observations at the NTT. JAS research was supported by a PhD fellowship of the 
International Max Planck Research School on Physical Processes in the Solar System and Beyond, and the Deutsche Forschungsgemeinschaft. VR research was supported by NASA 182 NEOO Program Grant 
NNX12AG12G, and NASA Planetary Geology and Geophysics Grant NNX11AN84G. We thank the IRTF TAC for awarding time to this project, and to the IRTF TOs and MKSS staff for their support. The authors would like to 
thank Paul Abell and an anonymous reviewer for their reviews, which helped to improve the manuscript.

\bibliographystyle{model2-names}
\bibliography{references}







\end{document}